\documentclass[aps,prd,twocolumn,superscriptaddress]{revtex4-2}

\usepackage[dvips]{graphicx}
\usepackage{ulem}
\usepackage[usenames]{color}
\usepackage{amsmath}
\usepackage{amssymb}
\usepackage{bm}
\usepackage{mathrsfs}
\usepackage{amsbsy}
\usepackage{hyperref}
\usepackage[all]{hypcap}
\usepackage{longtable}
\usepackage{url}
\usepackage{multirow}

\begin{document}

\title{Assignment of charmed-strange $D_{s0}(2590)^+$ and $D_{sJ}(3040)^+$}

\author{Zi-Han Jiang}

\author{Ailin Zhang}
\email{zhangal@shu.edu.cn}
\affiliation{Department of Physics, Shanghai University, Shanghai 200444, China}

\begin{abstract}
Based on analyses of the mass and the strong decay features, $D_{s0}(2590)^+$ observed by LHCb collaboration is identified as a radial excitation of the pseudoscalar $D_s$, and $D_{sJ}(3040)^+$ observed by BaBar collaboration is identified as a radial excitation of $D_{s1}(2536)^\pm$. $D_{s0}(2590)^+$ is possibly a pure $D_{s}(2~^1S_0)$ meson, both basic $D_{s1}(2536)^\pm$ and radially excited $D_{sJ}(3040)^+$ are possibly the mixtures $D_s(nP_1)$ between spin triplet $D_s(n~^3P_1)$ and spin singlet $D_s(n~^1P_1)$. In this arrangement, their masses meet the linear behavior of the radial Regge trajectory very well. In the $^3P_0$ strong decay model, the decay channels of $D_{s0}(2590)^+$ are $D^{*0}K^+$ and $D^{*+}K^0$, the total decay width is predicted with $\Gamma=76.12$ MeV. The main decay channels of $D_{sJ}(3040)^+$ are $D^{*0}K^+$/$D^{*+}K^0$ and $D^{*0}K^{*+}$/$D^{*+}K^{*0}$, the total decay width is predicted with $\Gamma=283.46$ MeV. These numerical strong decay results are consistent with the experiment data and support our arrangement. The dimensionless strength creation parameter $\gamma$ plays an important role in the calculation, and $\gamma=9.57$ is fixed through a comparison of the predicted strong decay widths of $D^*_{s2}(2573)$ and $D^*_{s3}(2860)^{\pm}$ with experimental data.
\end{abstract}

\maketitle

\section{Introduction}
A charmed-strange meson consists of a heavy charm quark/antiquark and a light strange antiquark/quark. The investigation of the mass spectrum, decays and productions of charmed-strange mesons will help people to understand the properties of quark dynamics and confinement inside mesons.

There are $11$ charmed-strange meson candidates listed in the Review of Particles Physics~\cite{pdg2022}. The orbital $S-$wave and $P-$wave charmed-strange excitations except for $D^*_{s0}(2317)^\pm$ and $D_{s1}(2460)^\pm$ have been established. Regarding $D^*_{s0}(2317)^\pm$ and $D_{s1}(2460)^\pm$, their masses are much lower than relevant theoretical predictions of $c\bar s$ mesons, and there exist many explanations of these two states outside the normal $c\bar s$ meson configuration. In recent year, more higher charmed-strange states were observed~\cite{pdg2022}. These observations arise people's great interest and provide more information on the quark structure and dynamics in mesons.

In $2021$, a new excited $D_s$ denoted as $D_{s0}(2590)^+$ in the $D^+K^+\pi^-$ final state was observed by the LHCb collaboration in $pp$ collision~\cite{lhcb2021}. Its mass and decay width have been measured with $M=2591\pm6\pm7$ MeV and $\Gamma=89\pm16\pm12$ MeV. Its spin parity was also determined to be $J^P=0^-$. The LHCb collaboration suggested it as a strong candidate for $D_{s}(2~^1S_0)$ state.

In the relativized Godfrey-Isgur quark model~\cite{prd32.189}, the predicted mass for $D_{s}(2~^1S_0)$ state is about $80$ MeV higher than the measured one of $D_{s0}(2590)^+$. To make the theoretical mass predictions consistent with experimental data, coupled channels calculations~\cite{prd104.094051,plb827.136998,prd106.074014,epjc83.1098} were proposed. In these coupled channels calculations, the interaction between meson pairs and mixing between $q\bar q$ meson and meson pair system were taken into account. In the coupled channels model~\cite{prd106.074014}, $D_{s0}(2590)^+$ is not thought a bare $c\bar s$ meson but a mixture with a large probability of meson pair system (coupled-channels). The mass of the mixed $D_{s}(2~^1S_0)$ was predicted with $2616$ MeV and its decay width was predicted with $112$ MeV in the $^3P_0$ model. $DK^*~(20.4\%)$ and $D^*K^*~(26.2\%)$ are its dominant coupling channels. In the same coupled channels model with a different potential~\cite{epjc83.1098}, the predicted $D_s(2~^1S_0)$ mass is about $50$ MeV higher than the mass of $D_{s0}(2590)^+$, while the predicted width $87$ MeV agrees well with the experimental data.

Instead of the coupled channels, an alternate modified potential model with screening effects~\cite{prd105.074037} was employed to calculate the mass spectrum of the charmed-strange mesons. The mass of pure $D_{s}(2~^1S_0)$ was predicted with $2620$ MeV and the strong decay width of $D_{s0}(2590)^+$ was predicted with $74.9$ MeV with the wave function obtained from a modified relativized quark model including the screening effects at $\gamma=9.32$ in the $^3P_0$ model. The partial decay widths to $D^{*+}K^0$ and $D^{*0}K^+$ are $35.5$ MeV and $39.4$ MeV, respectively.

Since the color hyperfine interaction, there is a mixing between $^3L_J$ meson and $^3L'_J$ meson. There is hence a mixing between $2~^3S_1$ and $1~^3D_1$ charmed mesons and charmed-strange mesons. $D^*_{s1}(2700)^\pm$ and $D^*_{s1}(2860)^\pm$ are popularly regarded as the mixtures of $2~^3S_1$ and $1~^3D_1$ $D_s$ states. As indicated in Ref.~\cite{prd102.054013}, the mass of the mixed radially excited state $D_{s1}^*$ is approximately the mass of the pure radially excited $2~^3S_1$ $D_s$ meson in a relativized quark model no matter how large the mixing is. A similar behavior exists in the charmonium system where the mixing of states with different $L$ can change the decay modes of the charmonium mesons, but cannot change dramatically the values of their masses~\cite{prd74.016002}. Though there are many explanations of $D^*_{s1}(2700)^\pm$, it is often identified with the radial excitation of $D_s^{*\pm}$ with $J^P=1^-$ apart from the mixing detail and quark dynamics.

$D_{sJ}(3040)^\pm$ was observed in inclusive production of $D^*K$ (not observed in the
$DK$ channel) in $e^+e^-$ annihilation by BaBar collaboration~\cite{babar2009}, this state was not observed in $D^{*0}K^+X$ final states in $pp$ collision by LHCb collaboration~\cite{lhcb2016}. As well known, there exists a mixing between the $~^3P_1$ and the $~^1P_1$ states via the spin orbit interaction or some other mechanism. In addition to some exotic explanations of $D_{s1}(2460)$, the two $J^P=1^+$ $D_s$ mesons ($D_{s1}(2460)$ and $D_{s1}(2536)$) are often regarded as the axial-vector mixing states between the $1~^3P_1$ and the $1~^1P_1$ $D_s$. Based on the analyses of mass and decay features, $D_{sJ}(3040)^+$ was popularly suggested as the mixing state between the $2~^3P_1$ and the $2~^1P_1$ $D_s$ with $J^P=1^+$.

However, there are different opinions on whose radial excitation $D_{sJ}(3040)^+$ is. $D_{sJ}(3040)^+$ was assigned as the first radial excitation of $D_s(2460)$ in Refs.~\cite{prd106.074014,prd80.071502,prd80.074037}, but was suggested as the first radial excitation of $D_{s1}(2536)$ in Ref.~\cite{epjc66.197}. Apart from the assignment of the basic $1P$ mixing $D_s$, $D_{sJ}(3040)^+$ was identified as the radially excited low-mass mixing state in Ref.~\cite{prd81.014031}, while was identified as the radially excited high-mass mixing state in Ref.~\cite{prd84.034006,prd97.054002}. Due to large theoretical and experimental uncertainties on the width, some analyses~\cite{epjc71.1582,prd86.054024,prd93.034035} indicated that $D_{sJ}(3040)^+$ may be identified with the radial excitation of $D_{s1}(2460)$ or the radial excitation of $D_{s1}(2536)$. That is to say, both identifications are possible. Outside the normal meson interpretation, $D_{sJ}(3040)^+$ was described as a $D^*(2600)K$ bound state~\cite{prd84.014013}.

It is observed that light $q\bar q$ mesons ($M<2400$ MeV) with the same $IJ^{PC}$ but different radial excitations form a Regge trajectory on $(n,~M^2)$ plots\cite{prd62.051502r}
\begin{equation}\label{regge}
M^2=M^2_0+(n-1)\mu^2
\end{equation}
where $n$ is the radial quantum numbers, $M_0$ is the mass of the basic meson and $\mu^2$ is the trajectory slope parameter which is approximately the same for all trajectories. This linear Regge trajectories feature was also observed in heavy quarkonia system~\cite{prd74.016002} and mentioned in $D_s$ mesons~\cite{prd80.071502}.

No matter whether $D_{s0}(2590)^+$, $D^*_{s1}(2700)^\pm$ and $D_{sJ}(3040)^+$ are pure $^{2S+1}L_J$ states or mixing states, they have interesting mass relations with $D_s$, $D^{*\pm}_s$ and $D_{s1}(2536)^\pm$, respectively
\begin{align}
M^2(D_{s0}(2590)^+)-M^2(D^\pm_s)=2.840~\mathrm{GeV^2},\nonumber  \\
M^2(D^*_{s1}(2700)^\pm)-M^2(D^{*\pm}_s)=2.905~\mathrm{GeV^2},\nonumber  \\
M^2(D_{sJ}(3040)^+)-M^2(D_{s1}(2536)^\pm)=2.840~\mathrm{GeV^2}.
\end{align}
Obviously, once $D_{s0}(2590)^+$, $D^*_{s1}(2700)^\pm$ and $D_{sJ}(3040)^+$ are identified with the radial excitations of $D_s$, $D^{*\pm}_s$ and $D_{s1}(2536)^\pm$, respectively, these relations meet Eq.~[\ref{regge}]. In other words, the masses of these radial $D_s$ mesons meet the linear Regge trajectories on the $(n,~M^2)$ plots very mell with a similar slope. This slope in $D_s$ system lies between the one in the light-meson system and the one in heavy quarkonia system.

The linear behavior of the Regge trajectories on $(n,~M^2)$ plane is a phenomenological observation of the experimental data in light-meson and heavy quarkonia systems, which indicates the 'string' nature of the mesons. The linear behavior of the Regge trajectories has also been explored and verified in some theoretical models~\cite{prd66.034025,prd66.034026,prd79.114029,jhep04(2004).039}. Though a $0^{-}$ radially excited $D_s$ denoted as $D'_s(2599)$ was already predicted ten years before the observation of $D_{s0}(2590)^+$ based only on a Regge trajectory analysis ~\cite{npa856.88}, whether there exists the linear behavior in heavy-light mesons has not been confirmed. Obviously, a test of the assignment of the radial excitations in other ways such as their strong decay features would provide an evidence.

$^3P_0$ strong decay model as an effective phenomenological method has been employed extensively and successfully to calculate the Okubo-Zweig-Iizuka(OZI)-allowed hadronic decay widths of mesons, baryons and even exotic multiquark states. In addition to the mass analysis, $^3P_0$ model was also employed to compute the strong decay widths of $D_{s0}(2590)^+$~\cite{prd104.094051,plb827.136998,prd106.074014,epjc83.1098,prd105.074037} and $D_{sJ}(3040)^+$~\cite{prd106.074014,prd80.074037,prd81.014031,epjc71.1582,prd93.034035}. However, the predictions of the strong decay widths are different for different assumptions of $D_{s0}(2590)^+$ and $D_{sJ}(3040)^+$. Further investigation of $D_{s0}(2590)^+$ and $D_{sJ}(3040)^+$ in the $^3P_0$ model is required.

The rest of this paper is organized as follows. The $^3P_0$ model is briefly introduced in Sec. 2. We will present the numerical results and analysis in Sec. 3. A summary is presented in the last section.
\section{A brief review of the $^3P_0$ model}
The phenomenological $^3P_0$ model is often known as the quark pair creation model. It was first proposed by Micu~\cite{npb10.521} and subsequently developed by Yaouanc {\it et al.}~\cite{prd8.2223,plb71.397}. The main idea of the $^3P_0$ model for a meson decay may be described in a simple picture as shown in Fig. 1. A pair of quark and antiquark in a flavor singlet with $J^{PC}=0^{++}$ (in a $^3P_0$ configuration) are assumed to be created from vacuum, and then regroup with the antiquark and the quark from an initial meson $A$ into final mesons $B$ and $C$.

\begin{figure}[ht]
\centering
\includegraphics[width=0.4\textwidth, angle=0]{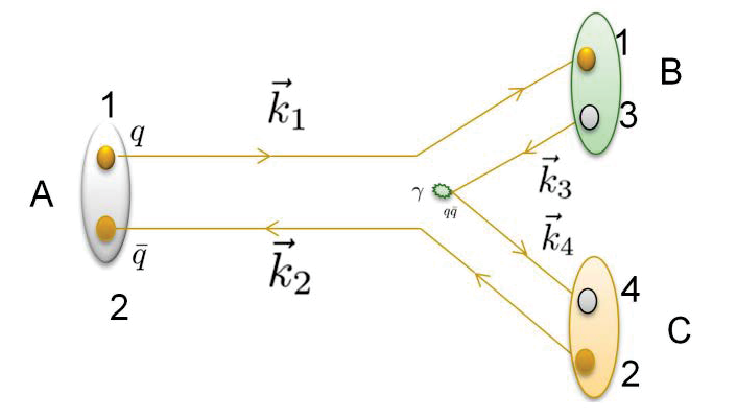}
\caption{Quarks arrangement of OZI-allowed strong decay $A\to B+C$\cite{prd94.114035}.}
\end{figure}

In the $^3P_0$ model, the decay width of $A\rightarrow B+C$ is
\begin{equation}\label{gamma}
\Gamma = \pi^2\frac{\mid \vec p \mid}{m^2_A}\sum_{JL}\vert \mathcal{M}^{JL}\vert^2
\end{equation}
where $\vec p$ is the momentum of mesons $B$ and $C$ in the initial meson $A$'s center-of-mass frame
\begin{equation}
\vert\vec p\vert=\frac{\sqrt{[m_A^2-(m_B-m_C)^2][m_A^2-(m_B+m_C)^2]}}{2m_A}.
\end{equation}
$\mathcal{M}^{JL}$ is the partial wave amplitude of $A\rightarrow B+C$ with $\vec J=\vec J_A+\vec J_B$, $\vec J_A=\vec J_B+\vec J_C+\vec L$ and $M_{J_A}=M_{J_B}+M_{J_C}$. In terms of the Jacob-Wick formula, one can derive the partial wave amplitude $\mathcal{M}^{JL}$ from the helicity amplitude $\mathcal{M}^{M_{J_A}M_{J_B}M_{J_C}}$
\begin{align}
\mathcal{M}^{JL}(A\rightarrow B+C)=&\frac{\sqrt{2L+1}}{2J_A+1}\sum_{M_{J_B}M{J_C}}\langle L0JM_{J_A}\vert J_AM_{J_A}\rangle \nonumber \\
                &\times\langle J_BM_{J_B}J_CM_{J_C}\vert JM_{J_A}\rangle\nonumber  \\
                &\times\mathcal{M}^{M_{J_A}M_{J_B}M_{J_C}}.
\end{align}
In the equation, the helicity amplitude is written as
\begin{align}\label{amplitude}
\mathcal{M}&^{M_{J_A}M_{J_B}M_{J_C}}\nonumber\\
=&\sqrt{8E_AE_BE_C}\gamma\sum_{\substack{M_{L_A},M_{S_A}\\M_{L_B},M_{S_B}\\M_{L_C},M_{S_C},m}}\langle L_AM_{L_A}S_AM_{S_A}\vert J_AM_{J_A}\rangle\nonumber\\
&\times\langle L_BM_{L_B}S_BM_{S_B}\vert J_BM_{J_B}\rangle\langle L_CM_{L_C}S_CM_{S_C}\vert J_CM_{J_C}\rangle\nonumber\\
&\times\langle1m;1-m\vert 00\rangle\langle\chi^{13}_{S_BM_{S_B}}\chi^{24}_{S_CM_{S_C}}\vert\chi^{12}_{S_AM_{S_A}}\chi^{34}_{1-m}\rangle\nonumber\\
&\times\langle\varphi^{13}_B\varphi^{24}_C\vert\varphi^{12}_A\varphi^{34}_0\rangle I^{M_{L_A},m}_{M_{L_B},M_{L_C}}(\vec p)
\end{align}
where $\gamma$ is a dimensionless parameter reflecting the strength of the quark/antiquark creation from vacuum and $I^{M_{L_A},m}_{M_{L_B},M_{L_C}}(\vec p)$ is the momentum integral
\begin{align}\label{integral}
I^{M_{L_A},m}_{M_{L_B},M_{L_C}}(\vec p)=&\int d\vec k_1 d\vec k_2d\vec k_3d\vec k_4 \nonumber\\
&\times\delta^3 (\vec k_1+\vec k_2-\vec p_A)\delta^3(\vec k_3+\vec k_4)\nonumber\\
&\times\delta^3(\vec p_B-\vec k_1-\vec k_3)\delta^3(\vec p_C-\vec k_2-\vec k_4)\nonumber\\
&\times\Psi^*_{n_BL_BM_{L_B}}(\vec k_{13})\Psi^*_{n_CL_CM_{L_C}}(\vec k_{24})\nonumber\\
&\times\Psi_{n_AL_AM_{L_A}}(\vec k_{12})y_{1m}(\vec k_{34}).
\end{align}
In Eq.~[\ref{integral}], $\vec k_{ij}=\frac{m_j\vec k_i-m_i\vec k_j}{\vec k_i+\vec k_j}$ is the relative momentum of quark $i$ and quark $j$. $y_{1m}(\vec k)$ denotes the solid harmonic polynomial corresponding to the quark pair.

Taking into account the symmetry of the total wave function (spacial, spin, flavor (isospin) and colour etc) of each meson, the flavor matrix element could be written as a product of the isospin matrix element with the Wigner $9j$ symbol after recoupling calculation~\cite{text,prd94.114035}
\begin{align}
\langle\varphi^{13}_B\varphi^{24}_C\vert\varphi^{12}_A\varphi^{34}_0\rangle
=&\sum_{I,I^3}\langle I_C I_C^3;I_B I_B^3\vert I_A, I_A^3\rangle\nonumber\\
&\times [(2I_B+1)(2I_C+1)(2I_A+1)]^{\frac{1}{2}}\nonumber\\
&\times
\begin{Bmatrix}
I_1&I_3&I_B\\
I_2&I_4&I_C\\
I_A&0&I_A
\end{Bmatrix}
\end{align}
where $I_i$ is the isospin of $u,~d,~s,~c$ quark, $I_A,~I_B,~I_C$ are the isospin of the mesons $A,~B$ and $C$. $I_A^3,~I_B^3,~I_C^3$ are the third isospin components of these three mesons. Isospin $I_P=0$ is assumed for the created quark pair.

Similarly, the spin matrix element is written as
\begin{align}
\langle\chi^{13}_{S_BM_{S_B}}&\chi^{24}_{S_CM_{S_C}}\vert\chi^{12}_{S_AM_{S_A}}\chi^{34}_{1-m}\rangle\nonumber\\
=&(-1)^{S_C+1}[3(2S_B+1)(2S_C+1)(2S_A+1)]^{\frac{1}{2}}\nonumber\\
&\times\sum_{S,M_S}\langle S_BM_{S_B}S_CM_{S_C}\vert SM_S\rangle\langle SM_S\vert S_AM_{S_A};1,-m\rangle\nonumber\\
&\times
\begin{Bmatrix}
1/2&1/2&S_B\\
1/2&1/2&S_C\\
S_A&1&S
\end{Bmatrix}
\end{align}
A simple harmonic oscillator(SHO) wave function is employed in our calculation as follows
\begin{align}
\Psi_{nLM_L}(\vec k)=&\frac{(-1)^n(-i)^L}{\beta^{3/2}}\sqrt{\frac{2n!}{\Gamma(n+L+3/2)}}(\frac{\vec k}{\beta})^L\nonumber\\
&\times \text{exp}(-\frac{\vec k^2}{2\beta^2})L^{L+1/2}_n(\frac{\vec k^2}{\beta^2})Y_{LM_L}(\Omega),
\end{align}
where $L^{L+1/2}_n(\frac{\vec k^2}{\beta^2})$ is the Lagueere polynomial function and $Y_{LM_L}(\Omega)$ is the spherical harmonic functions. $\beta$ is a harmonic oscillator dimensionless parameter. Further details of the equations, indices, matrix elements, and other indications for meson decay in the $^3P_0$ model can be found in Refs.~\cite{prd94.114035,epjc74.3031}.

\section{Strong decay of $D_{s0}(2590)^+$ and $D_{sJ}(3040)^+$}
\subsection{Input parameters in the $^3P_0$ model}

To proceed a practical calculation, we have to fix the parameters in the $^3P_0$ model. Masses of the constituent quarks are taken to be $m_d=m_u=220\, \text{MeV},\,m_s=419 \,\text{MeV}$ and $m_c=1628\,\text{MeV}$~\cite{prd93.034035}. Masses of mesons involved in the decays are chosen from the Review of Particles Physics~\cite{pdg2022}.

As well known, the dimensionless parameter $\gamma$ and the harmonic oscillator parameter $\beta$ are two important parameters in the $^3P_0$   model. $\gamma$ is usually regarded as a free constant fitted by experimental data. This $\gamma$ was sometimes not regarded as a constant in some applications in the $^3P_0$ model. For example, a relative momentum $p$ of the pair dependence of $\gamma$ was made use of in Refs.~\cite{fbs27.163,npa683.425}, and a scale with the mass of the pair dependence of the $\gamma$ was also investigated~\cite{plb715.322}.

The uncertainties resulted from $\beta$ have been investigated for a long time in the $^3P_0$ model~\cite{epjc74.3031,prd35.907,prd50.6855,arxiv1203.0370}. There are often two ways to fix the harmonic oscillator parameter $\beta$. One way is to fix it as a universal parameter~\cite{prd53.3700,prd68.054014,prd81.014021}. The other way is to fix it individually for each hadron~\cite{prd104.094051,prd93.034035,prd94.114035,epjc74.3031,prd73.054012,prd82.111501}. In this paper, a universal $\beta=400$ MeV is chosen for light mesons involved in the decay while individual $\beta$ are chosen for involved heavy-light mesons.

For the heavy-light mesons $D$ and $D_s$, individual effective values $\beta$ were obtained by reproducing the root mean square (rms) radius of the wave functions calculated in a relativized quark model through the harmonic oscillator wave function for the specified $(n,~l)$ quantum numbers~\cite{prd93.034035}. These $\beta$ values were employed in our calculation (see Table 1).
\begin{table}[h]\centering
\begin{tabular}{l c c c }
\hline
Mesons&$\beta$&Mesons&$\beta$\\
\hline
$D^{0(\pm)}$&601&$D^{*0(\pm)}$&516\\
$D_s$&651&$D^*_s$&562\\
$D^*_0(2300)^{0(\pm)}$&516&$D^*_2(2460)^{0(\pm)}$&437\\
$D_1(2430)^{0(\pm)}$&475~,~482&$D_1(2420)^{0(\pm)}$&475~,~482\\
$D_{s1}(2536)^{\pm}$&498~,~505&$D^*_{s2}(2573)$&464\\
$D_{s0}(2590)^+$&475&$D^*_{s3}(2860)^{\pm}$&426\\
$D^*_{sJ}(3040)$&433~,~434&{}&{}\\
\hline
\end{tabular}
\caption{Effective values of $\beta$ involved in our calculation (in MeV). If two $\beta$ are listed, the first one is for the spin singlet state and the second one is for the spin triplet state.}
\label{Table1}
\end{table}

From Eq.~(\ref{gamma}) and Eq.~(\ref{amplitude}), the partial decay width $\Gamma=\gamma^2\Gamma_0$, where the $\Gamma_0$ is the partial decay width with $\gamma=1$. Obviously, both the partial decay width and the total strong decay width are explicitly $\gamma^2$ dependent. Accordingly, the uncertainty resulted from parameter $\gamma$ is explicit. The branching fraction ratio is $\gamma$ independent.

Different $\gamma$ has been employed in the calculation of hadronic decay widths of mesons in different hadronic decay processes. $\gamma=6.25$ is employed in Refs.~\cite{arxiv1203.0370,prd53.3700,cpc37.023102}, $\gamma=6.95$ in Refs.~\cite{prd94.114035,prd93.034035} and $\gamma=8.7$ in Refs.~\cite{prd104.094051,prd91.054031,prd86.054025}.

In the coupling channel analyses of $D_{s0}(2590)^+$ in the $^3P_0$ model, the quark dependent effective pair creation strength $\gamma^{eff}_0={m_u\over m_i}\gamma$ was employed. $\gamma_0=0.529$~\cite{prd106.074014} and $\gamma_0=0.478$~\cite{epjc83.1098} were obtained by fitting the strong decay data of $D^*_{s2}(2573)$ ($1~^3P_2~D_s$), respectively. The $\gamma$ in these two literatures is $\frac{1}{\sqrt{96\pi}}$ as that in Refs.~\cite{prd104.094051,prd93.034035,prd94.114035,arxiv1203.0370,prd53.3700,cpc37.023102,prd91.054031,prd86.054025}, so $\gamma = 9.19$ and $\gamma = 8.30$ in our convention.

In the interpretation of $D_{s0}(2590)^+$ in the modified relativized model with screening effects, $\gamma=9.32$ was obtained from the strong decay behaviors of $D^*_{s2}(2573)$, $D^*_{s1}(2700)^\pm$, $D^*_{s1}(2860)^\pm$ and $D^*_{s3}(2860)^\pm$~\cite{prd105.074037}.

The dependence of decay widths on the parameter $\gamma$ is explicit, different choices of $\gamma$ may bring in large uncertainty (even to two times large) to the decay width. It is important to fix this parameter through suitable experimental data.

In fact, the nature of $D^*_{s1}(2700)^\pm$ and $D^*_{s1}(2860)^\pm$ is not definitely clear~\cite{prd102.054013}. In one hand, the detail of the mixing is not clear. $D^*_{s1}(2700)^\pm$ and $D^*_{s1}(2860)^\pm$ seem impossible to be identified as the mixtures of $2~^3S_1$ and $1~^3D_1$ $D_s$ mesons with a small mixing angle. On the other hand, the identification of $D^*_{s1}(2700)^\pm$ and $D^*_{s1}(2860)^\pm$ resolved from experimental data is not sufficient until now. It is not good to choose the experimental data of $D^*_{s1}(2700)^\pm$ and $D^*_{s1}(2860)^\pm$ to fix the parameter $\gamma$ in the $^3P_0$ model. Therefore, only the experimental data of $D^*_{s2}(2573)$ and $D^*_{s3}(2860)^\pm$ are used to fix the parameter $\gamma$ in our calcualtion.

Once $D^*_{s2}(2573)$ and $D^*_{s3}(2860)^{\pm}$ are identified with $1~^3P_2$ and $1~^3D_3$ $D_s$ meson, respectively, their decay channels are easy to be given out and corresponding hadronic decay widths could be calculated in the $^3P_0$ model. A most suitable $\gamma=9.57$ is determined by comparing theoretical results with experimental data, where a fitting process in Ref.~\cite{prd86.054025} is used. For a strange quark pair $s\bar s$ creation, $\gamma_{s\bar s}=\gamma/\sqrt 3$ is employed~\cite{plb71.397}. In Table 2, all the strong decay channels and relevant decay widths are presented at $\gamma=9.57$.
\begin{table}[h]\centering
\begin{tabular}{l c c c }
 \hline
  Mesons&Channels&Width($\gamma=9.57$)&Experiments\\
 \hline
 $D^*_{s2}(2573)$&$D^0K^+$&7.91&-\\
 ($1~^3P_2$)&$D^+K^0$&7.21&-\\
 {}&$D^{*0}K^+$&0.86&-\\
 {}&$D^{*+}K^0$&0.65&-\\
 {}&$D_s\eta$&0.04&-\\
 {}&${\Gamma(D^{*0}K^+)\over \Gamma(D^0K^+)}$&0.11&$<0.33$\\
 {}&Total&16.67&$16.9\pm 0.7$\\
 $D^*_{s3}(2860)^\pm$&$D^0K^+$&18.97&-\\
 ($1~^3D_3$)&$D^+K^0$&18.17&-\\
 {}&$D^{*0}K^+$&12.02&-\\
 {}&$D^{*+}K^0$&11.37&-\\
 {}&$D^0K^{*+}$&0.46&-\\
 {}&$D^+K^{*0}$&0.34&-\\
 {}&$D_s\eta$&0.63&-\\
 {}&$D_s^*\eta$&0.18&-\\
 {}&Total&62.14&$53\pm10$\\
 \hline
\end{tabular}
\caption{Decay channels and widths (in MeV) of $D^*_{s2}(2573)$ and $D^*_{s3}(2860)^\pm$ at $\gamma=9.57$.}
\label{Table2}
\end{table}

\subsection{Strong decay of $D_{s0}(2590)^+$}

As the analysis in the introduction, $D_{s0}(2590)^+$ may be assigned as the radial excitation of the pseudoscalar $D_s$. In the following, let's have a look at its strong decay feature as a pure $D_s(2~^1S_0)$ meson. In this assignment, $D_{s0}(2590)^+$ has $D^{*0}K^+$ and $D^{*+}K^0$ two decay channels. The hadronic decay widths are calculated at the fixed $\gamma=9.57$, and the results are presented in Table 3.

\begin{table}[h]\centering
\begin{tabular}{l c c c}
 \hline
Mesons&Channels& Width($\gamma=9.57)$ & Experiments\\
 \hline
 $D_{s0}(2590)^+$&$D^{*0}K^+$&40.03& -  \\
 $(2~^1S_0)$&$D^{*+}K^0$&36.09& - \\
 {}&$\frac{\Gamma(D^{*0}K^+)}{\Gamma(D^{*+}K^0)}$ &1.12& -\\
 {}&Total&76.12& $89\pm 16\pm12$\\
 \hline
\end{tabular}\\
\caption{Decay channels and widths (in MeV) of $D_{s0}(2590)^+$ at $\gamma=9.57$.}
\label{Table3}
\end{table}
According to the analysis in experiment~\cite{lhcb2021}, the measured total width of $D_{s0}(2590)^+$: $\Gamma=89\pm16\pm12$ MeV is described as the sum of two contributions. One is from the open decay channels to two-body $D^*K$ and the other is from three-body $DK\pi$ decays. The three-body decay contribution is usually very small. In the $^3P_0$ model, only two-body decay channels appear in the decay process of meson and the total hadronic decay width is the total decay width for two-body decays. Under theoretical and experimental uncertainties, the predicted total decay width ($\Gamma=76.12$ MeV) of $D_{s0}(2590)^+$ agrees with the experimental data very well. That is to say, the predicted hadronic decay width of $D_{s0}(2590)^+$ supports its identification as a pure $D_s(2~^1S_0)$ meson.

In order to have a reliable prediction, theoretical uncertainties have to be examined. As mentioned in Ref.~\cite{prd50.6855}, the explicit form of the meson wave functions has small impact on the final decay width, so the harmonic oscillator wave functions with the oscillator parameter $\beta$ is usually employed in the calculation. $\beta$ dependence of the decay width is implicit, where individual $\beta$ is employed for each heavy-light meson involved in the decay. In the following, the $\beta$ dependence of the decay widths of $D_{s0}(2590)^+$ is investigated. In a hadronic decay $A\to B+C$, $\beta_A$, $\beta_B$ and $\beta_C$ are denoted as the $\beta$ values of $D_{s0}(2590)^+$, charm meson and light meson $K$, respectively.

When $\beta_A$ and $\beta_B$ are fixed as those in Table 1, variations of the total width and $D^{*+}K^0$ partial width of $D_{s0}(2590)^+$ with $\beta_C$ are plotted from $300$ MeV to $500$ MeV at $\gamma=9.57$ in Figure 2(a). In the figure, a blue shaded region denotes the total width with uncertainties measured by LHCb and a blue dashed line indicates the central value of the total width. The predicted results agree well with the experiment in this range. The uncertainty from the variation of $\beta_C$ is not large, which is the reason that a universal $\beta_C=400$ MeV are chosen for all light mesons in our calculation.

As $\beta$ for all light mesons is fixed at $400$ MeV, the variations of the total width with $\beta_A$ and $\beta_B$ are calculated and shown in Figure 2(b). From this figure, the variation of $\beta_A$ and $\beta_B$ will bring in large uncertainty to the decay width. Therefore, it is very important to fix individual parameter $\beta$ properly for each heavy-light meson in the $^3P_0$ model.

\begin{figure*}[h]
\includegraphics[width=1.0\textwidth, angle=0]{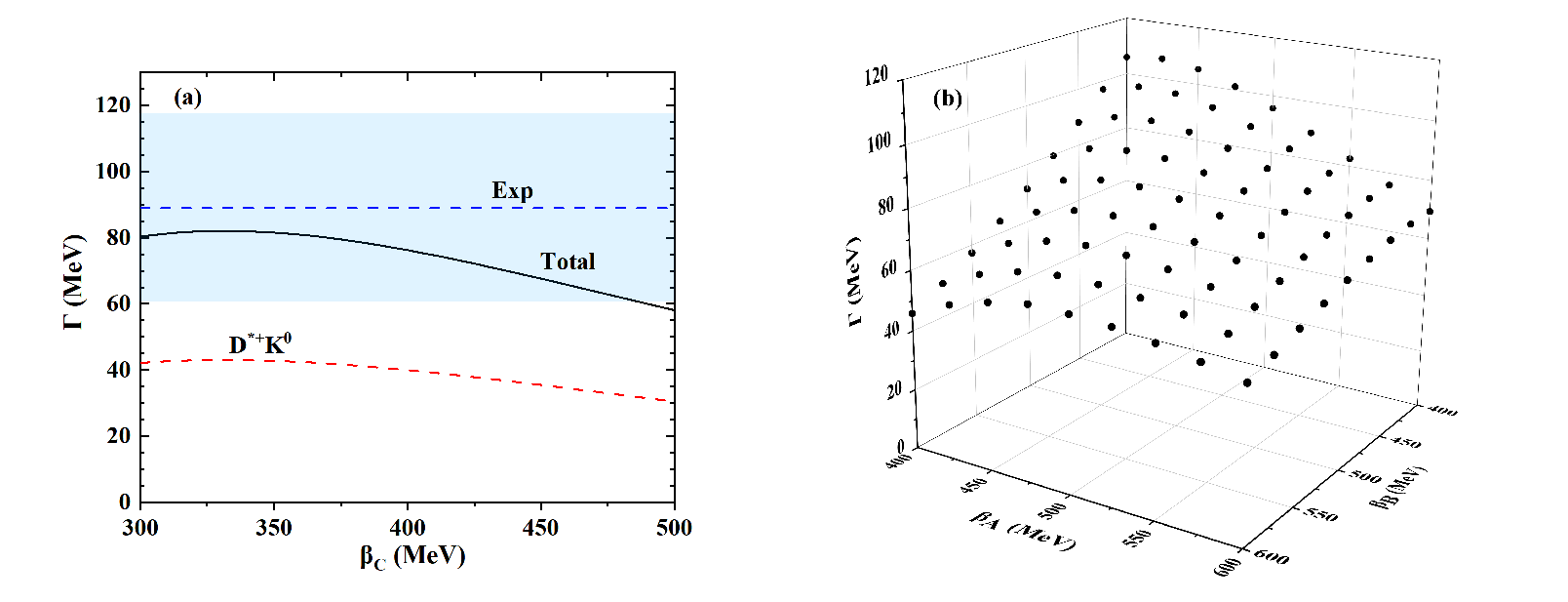}	
\caption{(a) Total width and $D^{*+}K^0$ partial width of $D_{s0}(2590)^+$ versus $\beta_C$ at $\gamma=9.57$. (b) Total width of $D_{s0}(2590)^+$ versus $\beta_A$ and $\beta_B$ at $\gamma=9.57$. The blue shaded region denotes the measured width (with errors) by LHCb. The blue dashed line indicates the central value}
	\label{fig_mom0}%
\end{figure*}

\subsection{Strong decay of $D_{s1}(2536)^\pm$ and $D_{sJ}(3040)^+$}

For Heavy-light $D_s$ mesons, charge conjugation parity is no longer a good quantum number. The mixing may occur between the spin singlet and the spin triplet via the spin orbit interaction. Therefore, $D_{s1}(2536)^\pm$ and $D_{sJ}(3040)^+$ are believed the mixtures of the $1P~D_s$ and $2P~D_s$, respectively. The detail of the mixture is~\cite{prd93.034035}
\begin{align}
    \begin{pmatrix}
    \vert nP_1\rangle\\
    \vert nP^{'}_1\rangle\\
    \end{pmatrix}
    = \begin{pmatrix}
    \cos{\theta_{nP}}&\sin{\theta_{nP}}\\
    -\sin{\theta_{nP}}&\cos{\theta_{nP}}\\
    \end{pmatrix}
     \begin{pmatrix}
    \vert n^1P_1\rangle\\
    \vert n^3P_1\rangle\\
    \end{pmatrix}\mbox{,}
\end{align}
where $|nP_1\rangle$ and $|nP'_1\rangle$ refer to the low-mass and high-mass mixing states between the spin singlet $|n^1P_1\rangle$ and the spin triplet $|n^3P_1\rangle$, respectively. The mixing angles $\theta_{nP}$ are chosen as $\theta_{1P}=-37.48^{\circ}$ for $n=1$, and $\theta_{2P}=-30.40^{\circ}$ for $n=2$~\cite{prd93.034035}.

In the following, two assignments, $nP_1$ and $nP'_1$, are tentatively assigned to $D_{s1}(2536)^+$ and $D_{sJ}(3040)^+$. Accordingly, possible decay channels and their strong decay widths are presented in Table 4 and Table 5.
\begin{table}[h]\centering
\begin{tabular}{l c c }
 \hline
 Channels&$D_{s1}(1P_1)$&$D_{s1}(1P'_1)$\\
 \hline
$D^{*0}K^+$&0.37&170.70\\
$D^{*+}K^0$&0.27&154.90\\
 Total&0.64&325.60\\
 Experiment&\multicolumn{2}{c}{$0.92\pm0.05$}\\
 \hline
\end{tabular}
\caption{Decay channels and widths (in MeV) of $D_{s1}(2536)^+$ in the two assignments at $\gamma=9.57$.}
\label{Table4}
\end{table}

\begin{table}[h]\centering
\begin{tabular}{l c c }
 \hline
 Channels&$D^*_{sJ}(2P_1)$&$D^*_{sJ}(2P'_1)$\\
 \hline
$D^{*0}K^+$&57.65&45.31\\
$D^{*+}K^0$&58.00&43.92\\
 $D^{0}K^{*+}$&11.5&33.4\\
 $D^{+}K^{*0}$&10.20&32.32\\
 $D^{*0}K^{*+}$&43.08&44.87\\
 $D^{*+}K^{*0}$&41.74&41.79\\
 $D^{*}_2(2460)^0K^+$&4.96&71.96\\
 $D^{*}_2(2460)^+K^0$&4.51&69.08\\
 $D^{*}_0(2300)^0K^+$&6.57&1.33\\
 $D^{*}_0(2300)^+K^0$&6.60&1.32\\
  $D_1(2420)^0K^+$&3.06&5.39\\
 $D_1(2420)^+K^0$&2.90&5.15\\
 $D_1(2430)^0K^+$&7.42&18.65\\
 $D_1(2430)^+K^0$&7.29&18.33\\
 $D^*_s\eta$&6.59&0.06\\
 $D_s\phi$&11.39&2.81\\
 Total&283.46&435.69\\
 Experiment&\multicolumn{2}{c}{$239\pm35^{+45}_{-42}$}\\
 \hline
\end{tabular}
\caption{Decay channels and widths (in MeV) of $D^*_{sJ}(3040)^+$ in two the assignments at $\gamma=9.57$.}
\label{Table5}
\end{table}

Comparing the theoretical results with experiment data, it seems like that $D_{s1}(2536)^+$ and $D^*_{sJ}(3040)$ are more possibly the mixing states $|nP_1\rangle$. In particular, their strong decay features support the assumption that $D^*_{sJ}(3040)$ is the radial excitation of $D_{s1}(2536)^+$.
From Table 5, the branching fraction ratio $\Gamma_{D^*K}/\Gamma_{D^*_2(2460)K}$ could be employed to distinguish the $|2P_1\rangle$ assignment and the $|2P'_1\rangle$ assignment of $D^*_{sJ}(3040)$ in experiment.

\section{Summary}
$D_{s0}(2590)^+$ was observed in the $D^+K^+\pi^-$ final state in $pp$ collision by LHCb collaboration. When its mass, decay width and spin-parity have been determined, $D_{s0}(2590)^+$ was suggested as a strong candidate for $D_{s}(2~^1S_0)$ state by LHCb collaboration. In existed literatures, the spectroscopy of $D_{s0}(2590)^+$ is explained by coupling-channels methods or modified quark potential model with screening effects.

$D_{sJ}(3040)^\pm$ was observed in inclusive production of $D^*K$ in $e^+e^-$ annihilation by BaBar collaboration but not observed in $DK$ channel. It was explained as the mixing axial-vector state between the $2~^3P_1$ and the $2~^1P_1$ states. However, people is not sure whether $D_{sJ}(3040)^\pm$ is the radial excitation of $D_{s1}(2460)$ or $D_{s1}(2536)^+$.

With a simple Regge trajectory analysis, we found that the masses of $D_{s0}(2590)^+$ and $D_{sJ}(3040)^\pm$ meet the radial linear trajectories on the $(n,~M^2)$ plane very well. In other words, $D_{s0}(2590)^+$ and $D_{sJ}(3040)^\pm$ are very possibly the radial excitations of the pseudoscalar $D_s$ and the axial-vector $D_{s1}(2536)^+$, respectively. Of course, the linear behavior of Regge trajectories in the heavy-light $D_s$ system requires more experimental data to be confirmed.

Under these assumptions, the strong decay features of $D_{s0}(2590)^+$ and $D_{sJ}(3040)^\pm$ are studied in the $^3P_0$ model. The numerical strong decay results support the assumption that $D_{s0}(2590)^+$ is the radial excitation of the pseudoscalar $D_s$. $D_{s0}(2590)^+$ is possibly the pure $D_s(2~^0S_1)$ meson. Both $D_{sJ}(3040)^\pm$ and $D_{s1}(2536)^+$ are very possibly the mixing states $|nP_1\rangle$ between the spin singlet $D_s(^1P_1)$ and the spin triplet $D_s(^3P_1)$. $D_{sJ}(3040)^\pm$ is the radial excitation of $D_{s1}(2536)^+$. In this assignment, $D_{s0}(2590)^+$ has $D^{*0}K^+$ and $D^{*+}K^0$ two decay channels. The total decay width of $D_{s0}(2590)^+$ is about $76.12$ MeV. Once the three-body decay contribution and uncertainties from theory and experiment have been taken into account, the predicted decay width agree with the experimental data very well. $D_{sJ}(3040)^+$ has main decay channels: $D^*K$ and $D^*K^*$. The total decay width of $D_{sJ}(3040)^+$ is about $283.46$ MeV. A forthcoming measurement of the branching fraction ratio $\Gamma_{D^*K}/\Gamma_{D^*_2(2460)K}$ will be important for the assignment of $D_{sJ}(3040)^+$.

In the calculation, the simple harmonic oscillator wave functions were employed for the mesons involved in the decays. Both the harmonic oscillator parameters $\beta$ and the dimensionless quark/antiquark creation strength parameter $\gamma$ play important roles in the calculation of decay width. The decay width is explicitly $\gamma^2$ dependent, while the dependence of the decay width on $\beta$ is implicit. With the experimental data of $D^*_{s2}(2573)$ and $D^*_{s3}(2860)^{\pm}$ which are identified with $1~^3P_2$ and $1~^3D_3$ $D_s$ meson, respectively, the $\gamma$ is fixed at $9.57$. The uncertainty from $\beta$ and $\gamma$ will bring in some uncertainties to the decay widths. In order to obtain a more accurate decay width in the $^3P_0$ model, it is important to fix these parameters both in theory and in experiment.

\section*{Acknowledgements}
This work is supported by National Natural Science Foundation of China under the Grant No. 11975146.

\end{document}